\documentclass[twoside,twocolumn,english,prc,amsmath,a4paper,myheadings]{revtex4}
\usepackage[T1]{fontenc}
\usepackage{geometry}
\usepackage{amsmath}
\usepackage{graphicx}
\usepackage{amssymb}

\makeatletter
\def\@oddhead{\rightmark \hfill  The "Ridge" in Proton-Proton Scattering at 7 TeV  \hfill \thepage}
\def\@evenhead{\thepage \hfill K. Werner et al.\hfill}
\def\fnum@table{\tablename~{\bf\thetable}}
\def\fnum@figure{\figurename~{\bf\thefigure}}
\def\tablename{\footnotesize{\bf Table}}
\def\figurename{\footnotesize{\bf Figure}}

\usepackage{dcolumn}

\def\citet{\cite}

\makeatother

\usepackage{babel}

\begin{document}

\title{The {}``Ridge'' in Proton-Proton Scattering at 7 TeV}

\author{{\normalsize K.$\,$Werner$^{(a)}$, Iu.$\,$Karpenko$^{(b)}$, T.$\,$Pierog$^{(c)}$}}

\address{$^{(a)}$ SUBATECH, University of Nantes -- IN2P3/CNRS-- EMN, Nantes,
France}

\address{$^{(b)}$ Bogolyubov Institute for Theoretical Physics, Kiev 143,
03680, Ukraine}

\address{$^{(c)}$Karlsruhe Institute of Technology (KIT) - Campus North,
Institut f. Kernphysik, Germany}

\begin{abstract}
One of the most important experimental results for proton-proton scattering
at the LHC is the observation of a so-called {}``ridge'' structure
in the two particle correlation function versus the pseudorapidity
difference $\Delta\eta$ and the azimuthal angle difference $\Delta\phi$.
One finds a strong correlation around $\Delta\phi=0$, extended over
many units in $\Delta\eta$. We show that a hydrodynamical expansion
based on flux tube initial conditions leads in a natural way to the
observed structure. To get this result, we have to perform an event-by-event
calculation, because the effect is due to statistical fluctuations
of the initial conditions, together with a subsequent collective expansion.
This is a strong point in favour of a fluid-like behavior even in
$pp$ scattering, where we have to deal with length scales of the
order of  $0.1$ fm.
\end{abstract}
\maketitle
The CMS collaboration published recently results \citet{cms_ridge}
on two particle correlations in $\Delta\eta$ and $\Delta\phi$, in
$pp$ scattering at 7 TeV. Most remarkable is the discovery of a ridge-like
structure around $\Delta\phi=0$, extended over many units in $\Delta\eta$,
referred to as {}``the ridge'', in high multiplicity $pp$ events.
A similar structure has been observed in heavy ion collisions at RHIC,
and there is little doubt that the phenomenon is related to the hydrodynamical
evolution of matter \citet{intro1,intro2,intro3,intro4}. This {}``fluid
dynamical behavior'' is actually considered %
\begin{figure}[b]
\begin{centering}
\includegraphics[scale=0.45]{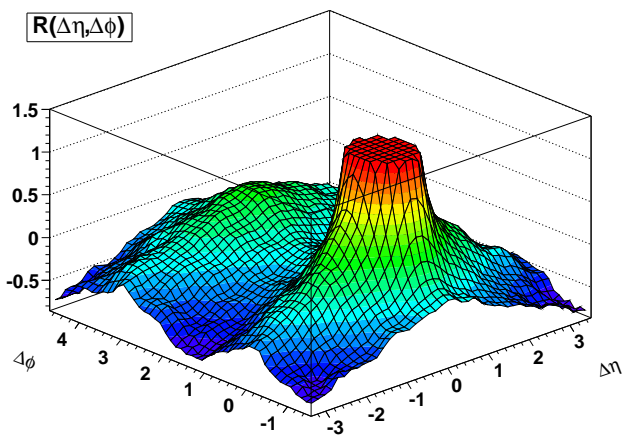}
\par\end{centering}

\caption{(Color online) Two particle correlation function $R$ versus $\Delta\eta$
and $\Delta\phi$ for high multiplicity events in $pp$ collisions
at 7 TeV, as obtained from a hydrodynamical evolution based on flux
tube initial conditions. We consider particles with $p_{t}$ between
$1$ and 3 GeV/c. \label{cap:ridge} }

\end{figure}
to be the major discovery at RHIC mainly based on the studies of azimuthal
anisotropies \citet{hydro1,hydro1b,hydro1c,hydro1d,hydro1e}.

\begin{figure}[b]
\begin{centering}
\includegraphics[scale=0.45]{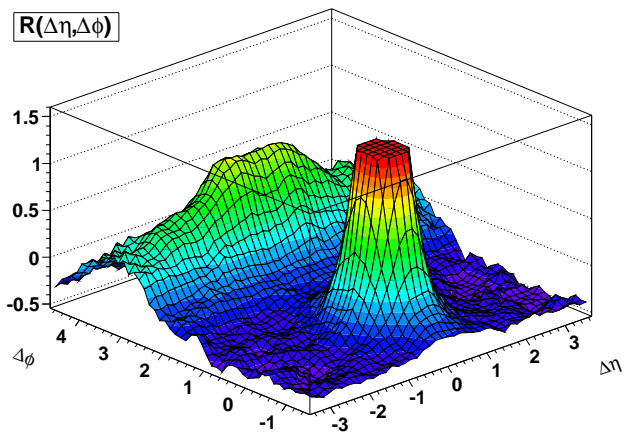}
\par\end{centering}

\caption{(Color online) Same as figure \ref{cap:ridge}, but calculation without
hydro evolution .i.e. particle production directly from string (flux
tube) decay. \label{cap:ridge2} }

\end{figure}

So does $pp$ scattering provide as well a liquid, just ten times
smaller than a heavy ion collision? It seems so! We showed recently
\citet{epos2pp} that if we take exactly the same hydrodynamic approach
which has been so successful for heavy ion collisions at RHIC \citet{epos2},
and apply it to $pp$ scattering, we obtain already very encouraging
results compared to $pp$ data at 0.9 TeV. In this paper, we apply
this fluid approach, always the same procedure, to understand the
7 TeV results. Before discussing the details of the approach, we present
the most important results of this work, namely the correlation function.
In fig. \ref{cap:ridge}, we show that our hydrodynamic picture indeed
leads to a near-side ridge, around $\Delta\phi=0$, extended over
many units in $\Delta\eta$. In fig. \ref{cap:ridge2},%
\begin{figure*}[tb]
\begin{centering}
\includegraphics[scale=0.33]{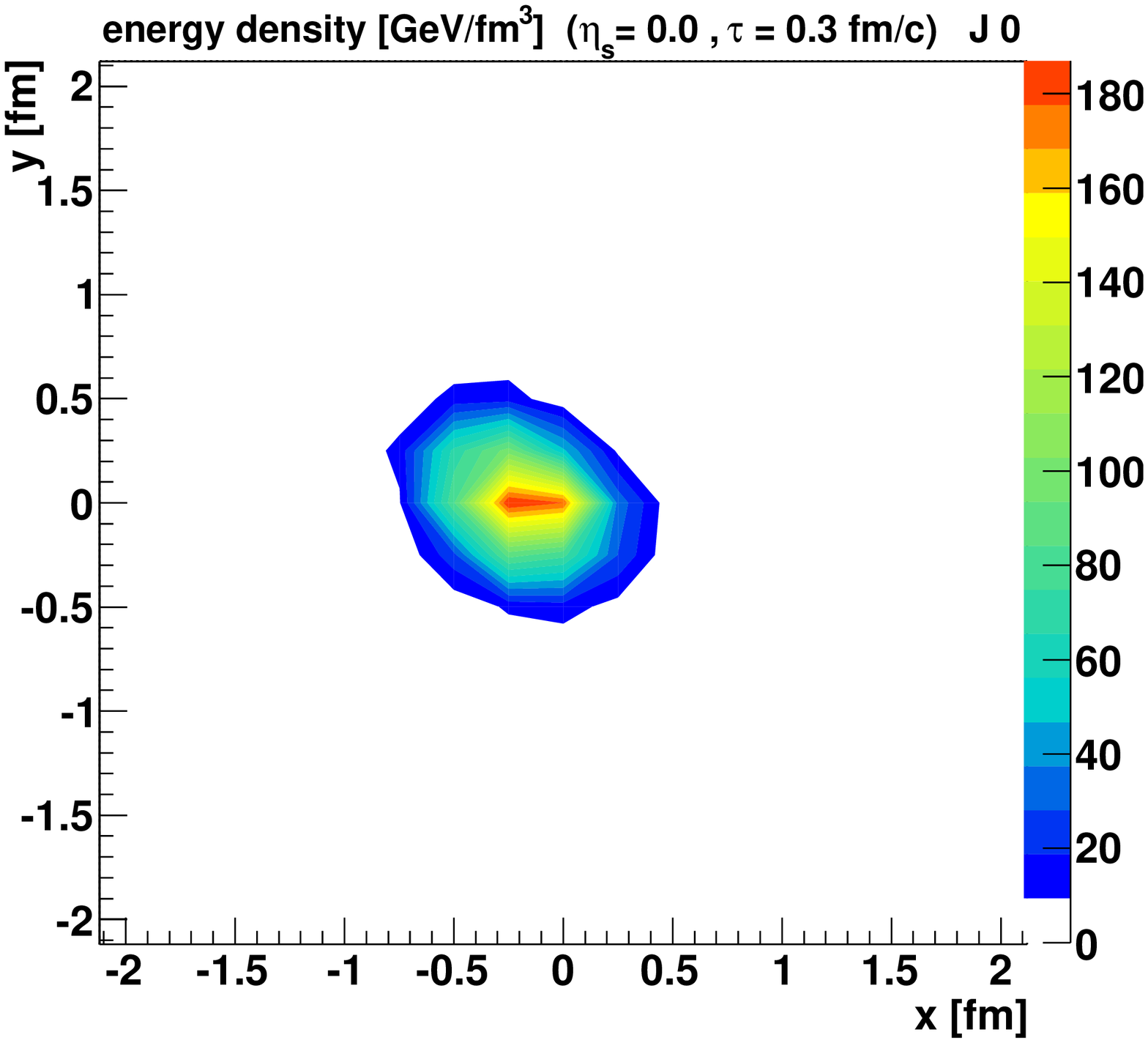}\includegraphics[scale=0.33]{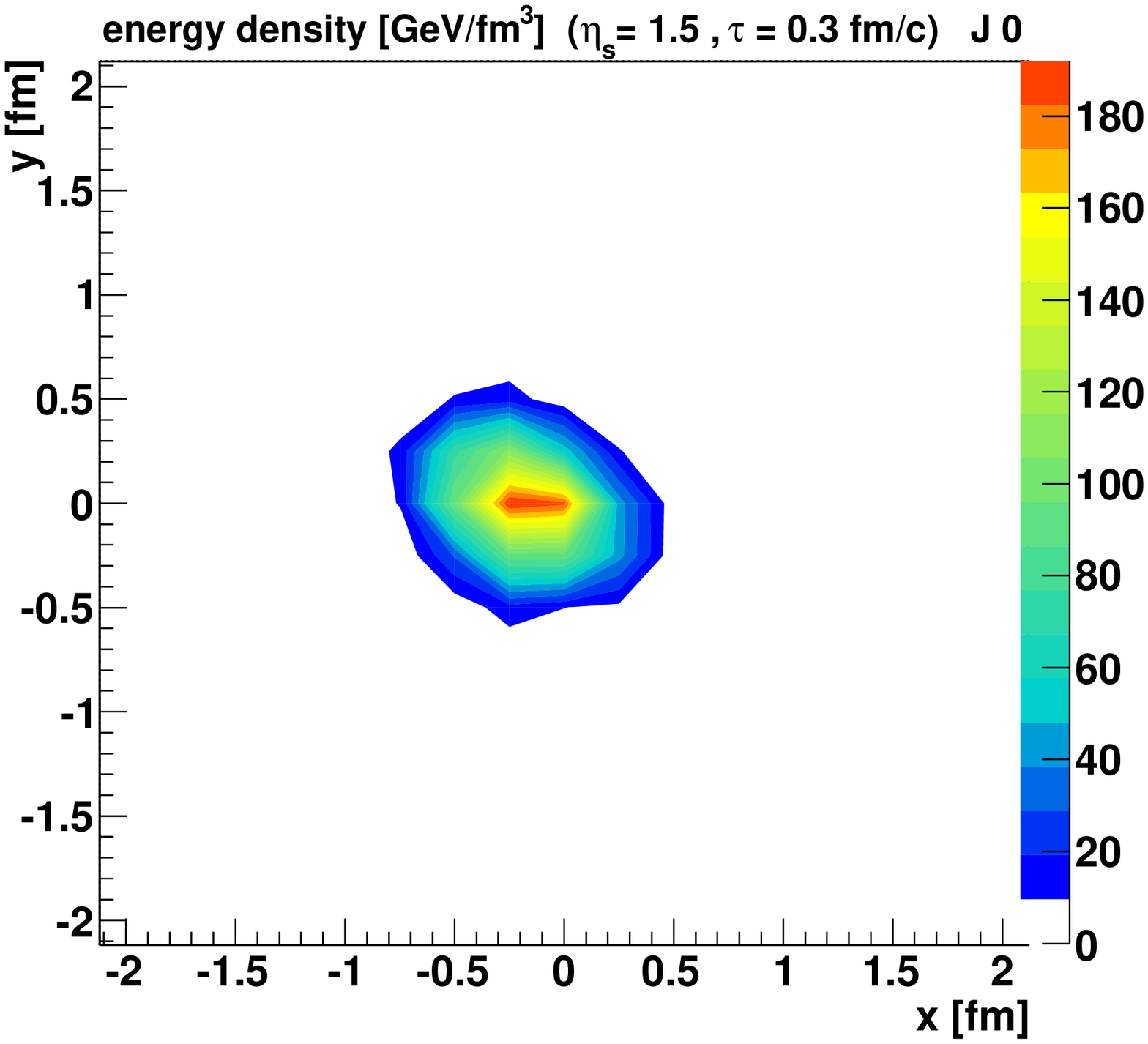}\\
\includegraphics[scale=0.33]{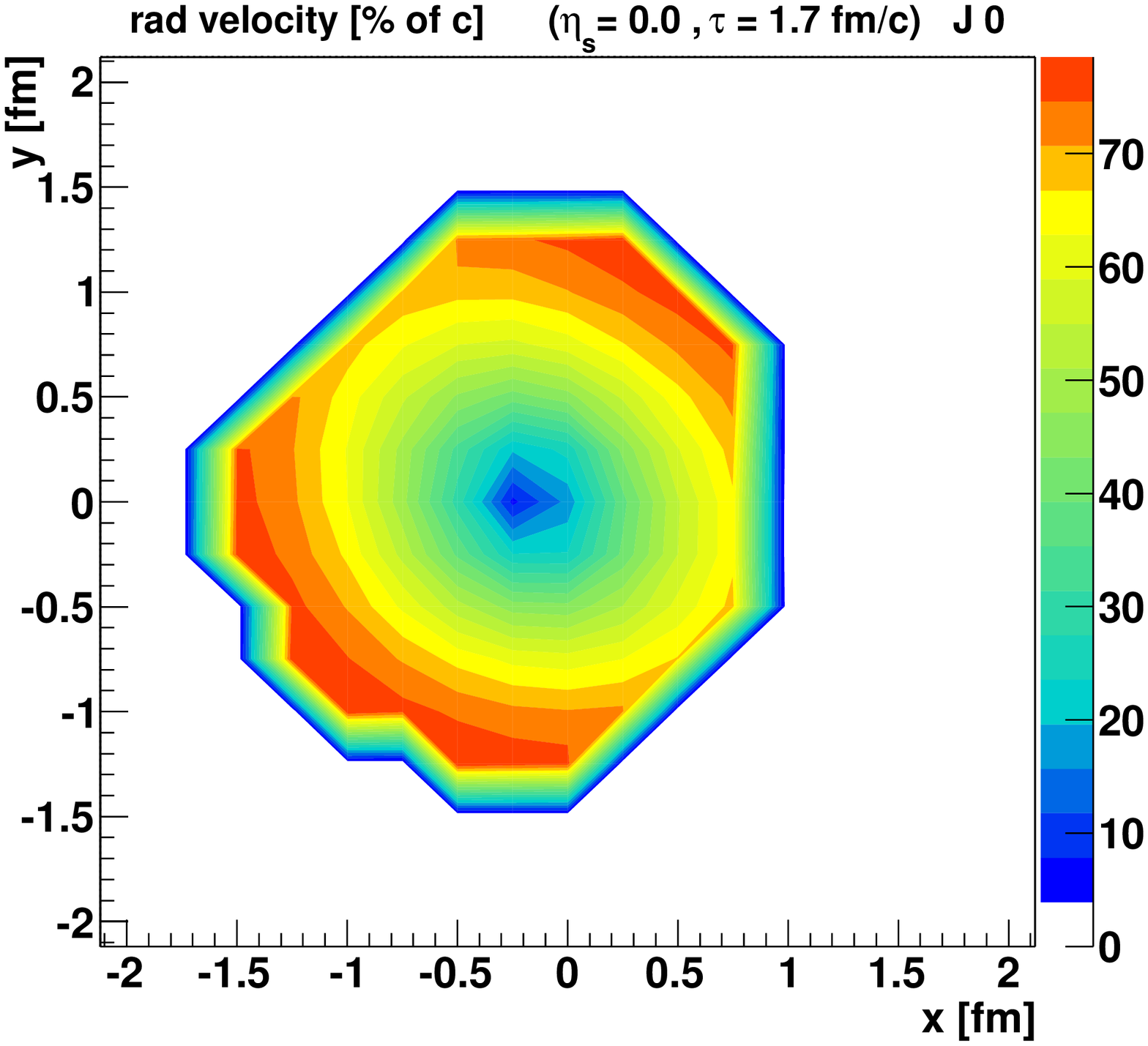}\includegraphics[scale=0.33]{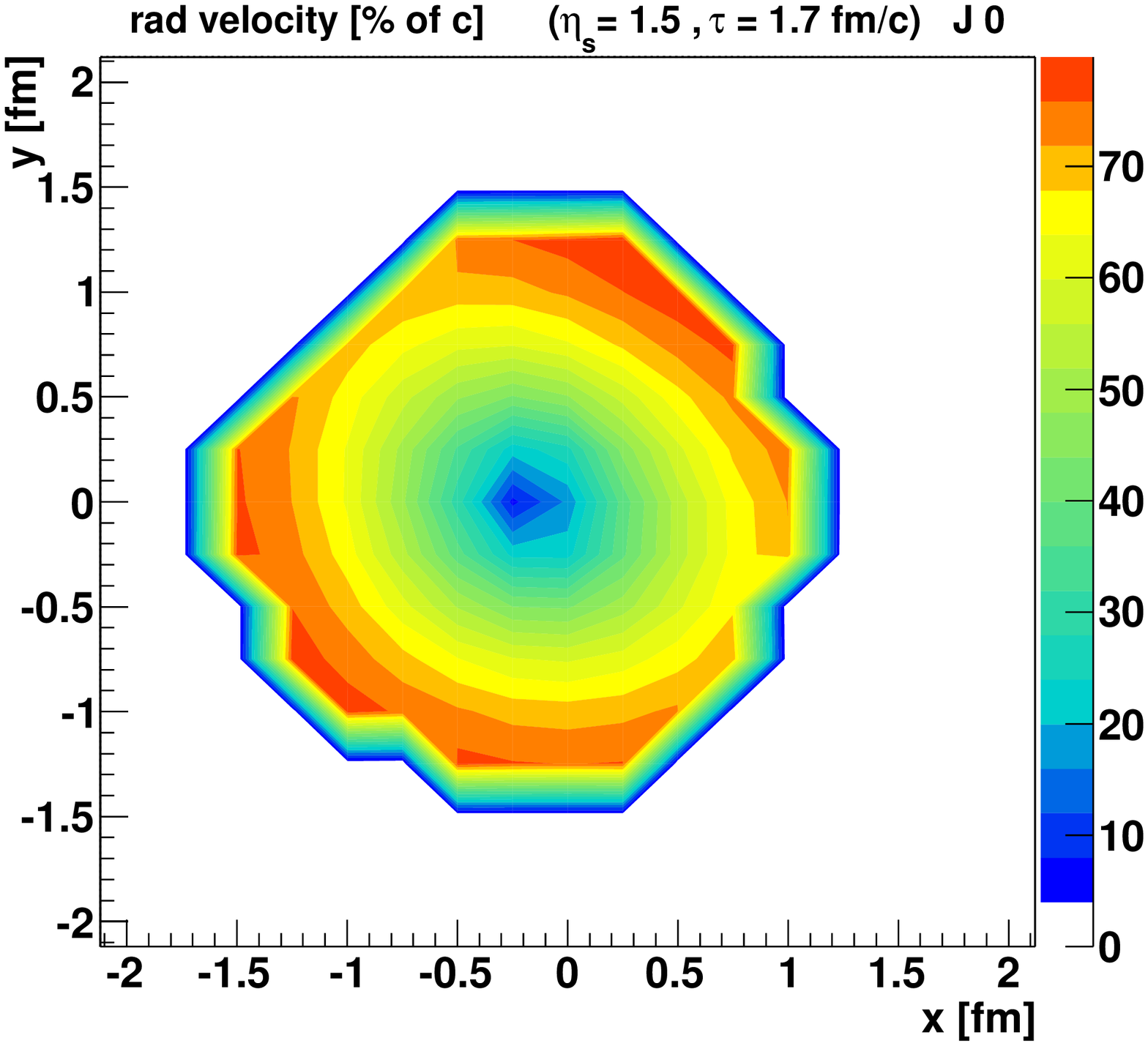}
\par\end{centering}

\caption{(Color online) Initial energy density (upper panel) and radial flow
velocity at a later time (lower panel) for a high multiplicity $pp$
collision at 7 TeV at a space-time rapidity $\eta_{s}=0$ (left) and
$\eta_{s}=1.5$ (right). \label{cap:ei}}

\end{figure*}
we show in the corresponding result for the pure basic string model,
without hydro evolution. There is no ridge any more! This shows that
the hydrodynamical evolution {}``makes'' the effect. One should
note that the correlation functions are defined and normalized as
in the CMS publication, so we can say that our {}``ridge'' is quite
close in shape and in magnitude compared the experimental result.
The experimental high multiplicity bin corresponds to about 7 times
average, whereas in our calculation (extremely demanding concerning
CPU power) {}``high multiplicity'' refers to 5.3 times average (we
actually trigger on events with 10 elementary scatterings). We cannot
go beyond at the moment.

It is easy to understand the origin of the ridge, in a hydrodynamical
approach based on flux tube initial conditions. Imagine many (say
20) flux tubes of small transverse size (radius $\approx0.2$ fm),
but very long (many units of space-time rapidity $\eta_{s}$ ). For
a given event, their transverse positions are randomly distributed
within the overlap area of the two protons. Even for zero impact parameter
(which dominated for high multiplicity events), this randomness produces
azimuthal asymmetries, as shown in fig. \ref{cap:ei}, upper panel.
The energy density obtained from the overlapping flux tubes (details
will be discussed later) shows an elliptical shape. And since the
flux tubes are long, and only the transverse positions are random,
we observe the same asymmetry at different longitudinal positions
($\eta=0$ and $\eta=1.5$ in the figure). So we observe a translational
invariant azimuthal asymmetry! 

If one takes this asymmetric but translational invariant energy density
as initial condition for a hydrodynamical evolution, the translational
invariance is conserved, and in particular translated into other quantities,
like the flow. In fig. \ref{cap:ei}, lower panel, we show the radial
flow velocity at a later time again at the two space-time rapidities
$\eta_{s}=0$ (left) and $\eta_{s}=1.5$ (right). In both cases, the
flow is more developed along the direction perpendicular to the principal
axis of the initial energy density ellipse. This is a very typical
fluid dynamical phenomenon, referred to as elliptical flow. 

Finally, particles are produced from the flowing liquid, with a preference
in the direction of large flow. This preferred direction is therefore
the same at different values of $\eta_{s}$. And since $\eta_{s}$
and pseudorapidity $\eta$ are highly correlated, one observes a $\Delta\eta$$\Delta\phi$
correlation, around $\Delta\eta=0$, extended over many units in $\Delta\eta$:
a particle emitted a some pseudorapidity $\eta$ has a large chance
to see a second particle at any pseudorapidity to be emitted in the
same azimuthal direction.

Here, a couple of remarks are in order. It should be mentioned that
the magnitude of the radial flow (and all observables affected by
this flow) are depending on the choice of the flux tube radius. A
bigger radius leads to smaller flow. The value of 0.2 fm has been
chosen to get an overall best picture for all observables depending
on flow. In our picture, the ridge effect is biggest at intermediate
values of $p_{t}$, because at lower $p_{t}$ the effect from flow
on the particle $p_{t}$ is small (flow can only increase $p_{t}$),
whereas at large $p_{t}$ the effect has to disappear because particles
are coming from jets rather than the fragmenting fluid. Finally we
have to admit that -- although the ridge seems to be reproduced in
form a magnitude -- the awayside ridge is too low in the simulation.
The problem with the awayside region is the fact that here the cut
between core and corona is crucial. In a first version we allowed
all string segments with $p_{t}$ larger than 3 GeV/c to escape from
the core, with the result of having almost no awayside correlation,
because all candidates were included in the core plasma. So we were
forced to reduce this cutoff to 1 GeV/c, which gives some awayside
ridge without destroying the $p_{t}$ spectra. In reality there is
of course no cutoff but {}``some continuous procedure'', but this
is a project for the future. It should also be mentioned that momentum
conservation contributes to the correlation, as discussed in \citet{bozek},
and the usual hadronization procedure in hydrodynamical calculations
(Cooper-Frye) does not conserve momentum event-by-event. However,
this should not modify the form of the near-side ridge. In fig. \ref{cap:ridge2},
we show a calculation with perfect momentum conservation, and the
effect on the near-side is indeed a reduction of the correlation function
(negative values), but its form is a plateau, not a valley.

\begin{figure}[b]
\begin{centering}
{\Large (a)~}\includegraphics[angle=270,scale=0.3]{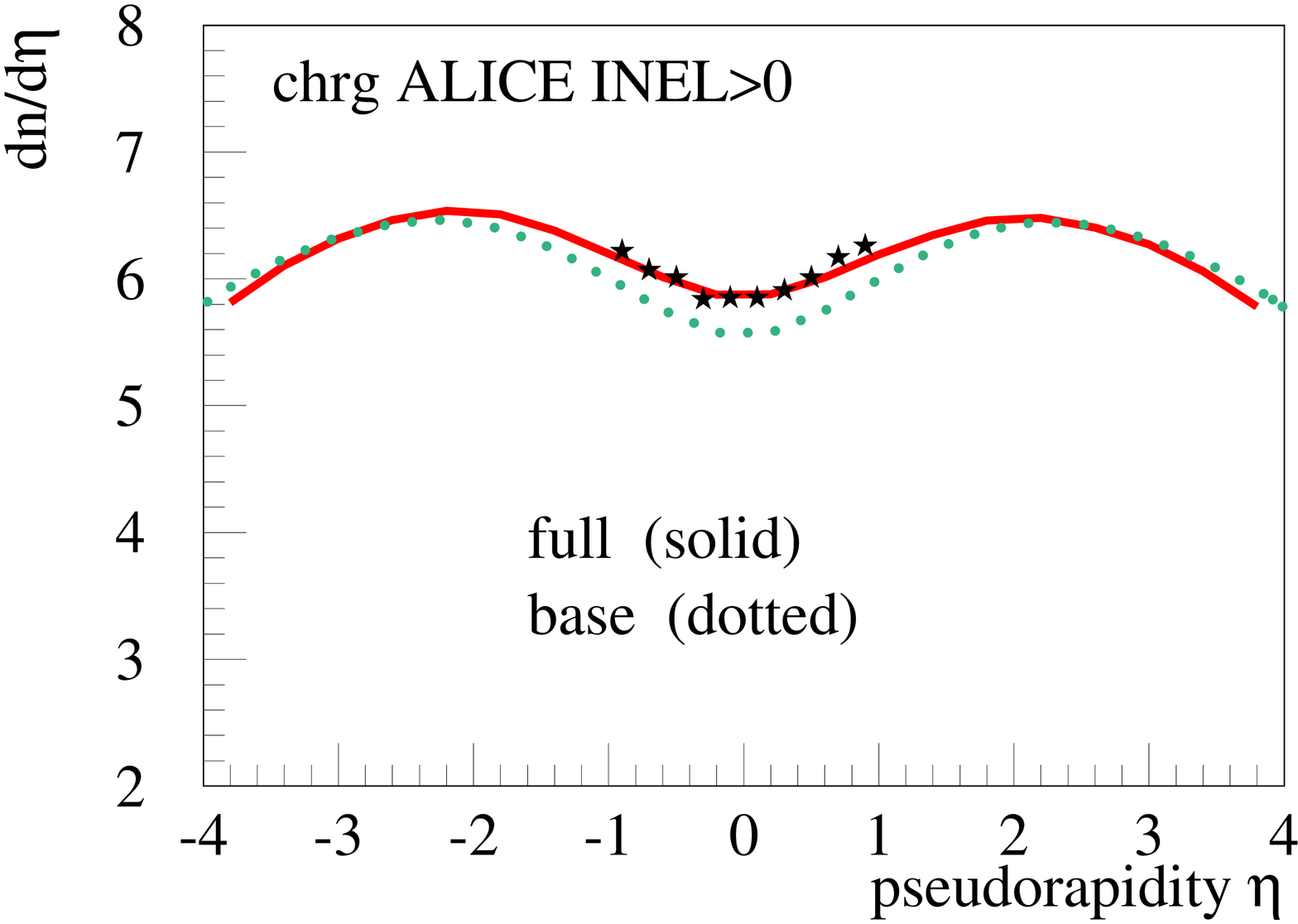}
\par\end{centering}

\vspace{-0.7cm}

\begin{centering}
{\Large (b)~}\includegraphics[angle=270,scale=0.3]{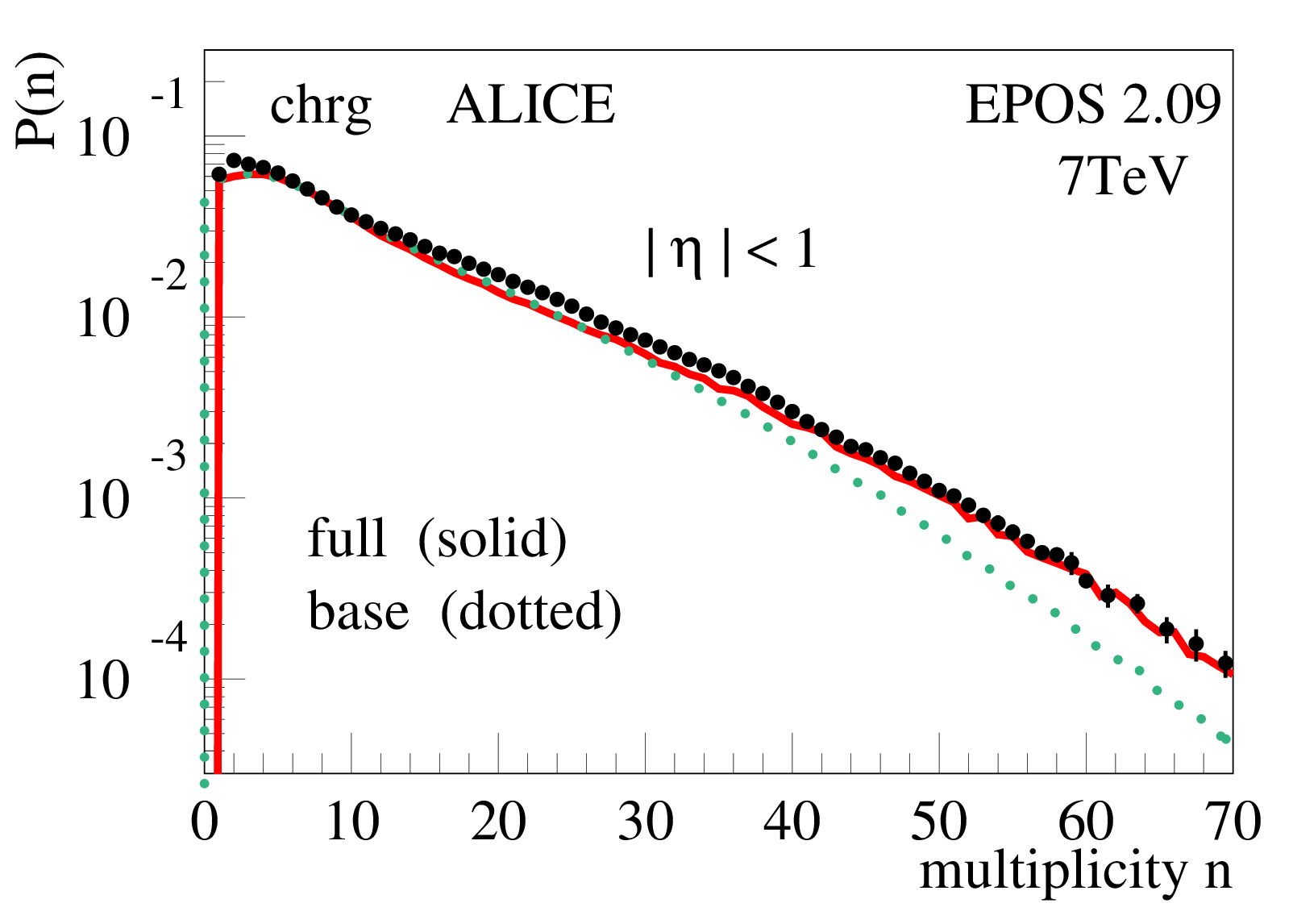}\vspace{-0.7cm}

\par\end{centering}

\begin{centering}
{\Large (c)~}\includegraphics[angle=270,scale=0.3]{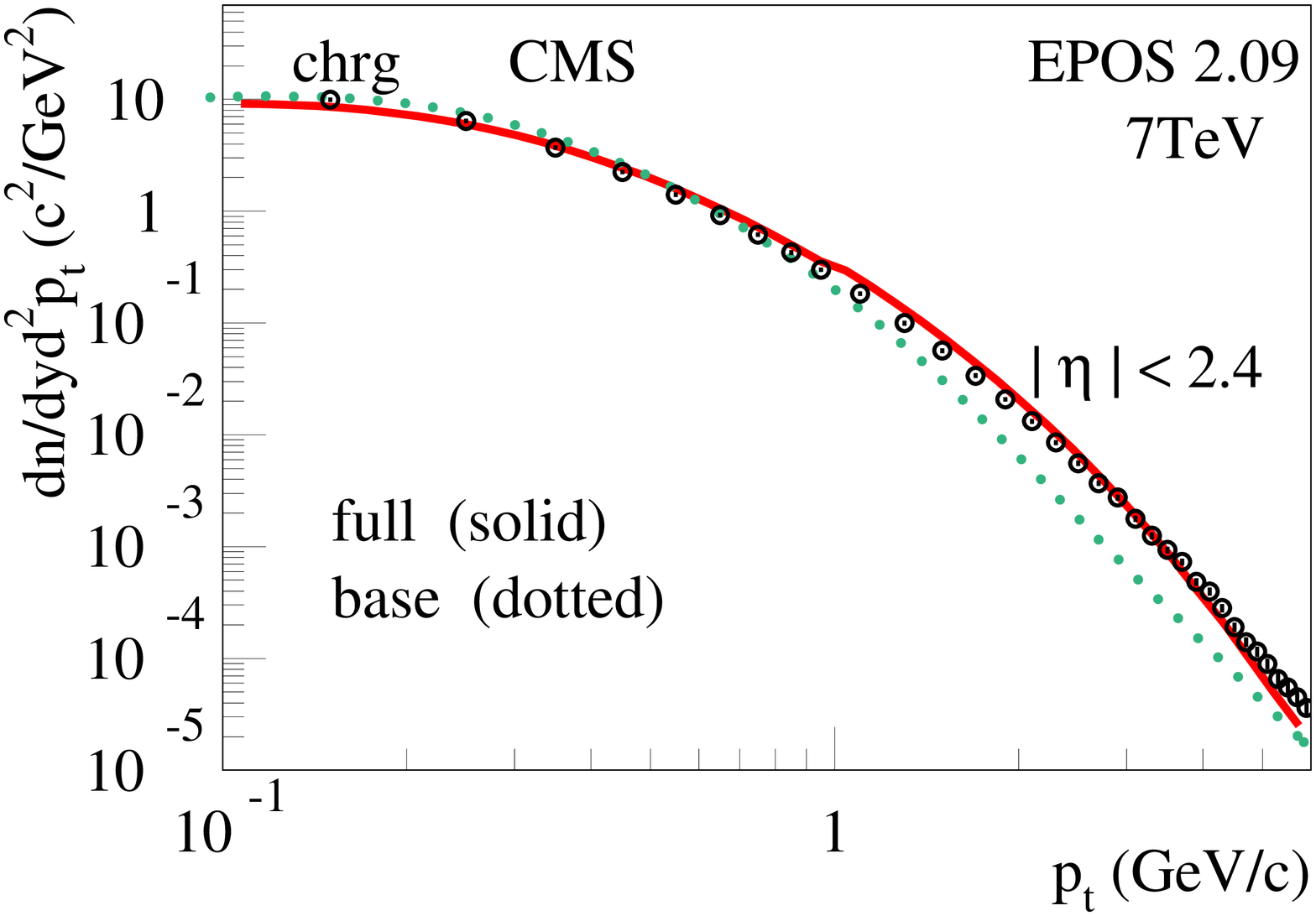}\\

\par\end{centering}

\caption{(Color online) Pseudorapidity distribution (INEL>0 trigger) (a), multiplicity
distribution (b), and transverse momentum distribution (c)  in $pp$
scattering at 7 TeV, compared to data (points). We show the full calculation
(solid line), and a calculation without hydrodynamic evolution (dotted).\label{fig:element}}

\end{figure}

In our approach the elliptical flow plays an important role, as discussed
earlier. This is perfectly compatible with a recent analysis \citet{bozek},
where the ridge correlation is obtained from a elliptical parametrization
of particle spectra. The are a couple of publications discussing elliptical
flow in $pp$. Closest to our approach is the work presented in \citet{solana},
where the eccentricity $\epsilon$ in $pp$ scattering is obtained
from statistical fluctuations, as in our model. The elliptical flow
$v_{2}$ is then obtained simply from an empirical $v_{2}/\epsilon$
relation. In refs. \citet{david,prasad}, an initial eccentricity
$\epsilon$ is obtained from a Glauber type model similar to the one
employed for heavy ion collisions, which leads to elliptical flow
using 2D hydrodynamics \citet{david} or an empirical $v_{2}/\epsilon$
relation \citet{prasad}. The Glauber picture is quite different to
ours, where the main origin of asymmetry are statistical fluctuations,
not geometry. Finally, also in \citet{kodama}, elliptical flow is
obtained in a hydrodynamical calculation, but here based on parametrized
initial conditions. We obtain a numerical value for the integrated
$v_{2}$ of about 0.01 at midrapidity, compatibel with values of about
0.05 at intermediate pt from other calculations \citet{pajares,bozek}.

Elliptical flow is an important issue, but crucial for the discussion
in this paper is, however, the fact that the (elliptical) asymmetry
of the flow is translational invariant, coming from the flux tube
structure. So the main point of this paper is not the elliptical flow
itself, but the fact that it is translational invariant, which leads
to the long range structure. Another important issue is the randomness
of the initial conditions. To treat all these elements in a realistic
calculation, we present here for the first time an even-by-event treatment
(see also \citet{hydro4,epos2}) of the 3+1 dimensional hydrodynamical
evolution for $pp$ scattering, based on random initial conditions.
This is an enormous computational effort. A $pp$ calculation is as
demanding as a heavy ion scattering: the volume is smaller, but the
cell sizes as well. On the other hand the multiplicities in $pp$
are small, so in particular for correlation studies a very large number
of events has to be simulated. Triggering is much more difficult in
$pp$ compared to $AA$, because in $pp$ multiplicity and geometrical
centrality are much less correlated than in $AA$.

Our hydrodynamical approach gives {}``a natural explanation'' of
the ridge phenomenon, without any need to construct asymmetries by
hand. This is a strong point in favor of a collective fluid-like behavior
of matter even in $pp$ scattering, which is still considered by many
people as an {}``elementary interaction''. 

The {}``flux tube + hydro'' approach has been extensively discussed
in \citet{epos2pp,epos2}. Crucial is an event-by-event treatment
of the hydrodynamic evolution (3D treatment, realistic equation of
state), where the initial condition for each event is obtained from
an EPOS 2 calculation. This is a multiple scattering approach, providing
multiple {}``parton ladders'', which are identified with elementary
flux tubes \citet{cgc}, the latter ones treated as classical strings.
In case of very high energy proton-proton collisions, in particular
for large numbers of scatterings, in a large fraction of the volume
the density of strings will be so high that they cannot possibly decay
independently. Instead, based on the four-momenta of infinitesimal
string segments, one computes the energy density $\varepsilon(\tau_{0},\vec{x})$
(see fig. \ref{cap:ei}) and the flow velocity $\vec{v}(\tau_{0},\vec{x})$,
which serve as initial conditions for the subsequent hydrodynamic
evolution, which lets the system expand and cool down till freeze
out at some $T_{H}$ according to the Cooper-Frye prescription.

Our above-mentioned results concerning the ridge are only meaningful
if the model can reproduce elementary distributions. In the following
we will compare two different scenarios: the full calculations, including
hydro evolution (full), and a calculation without hydrodynamical evolution
(base). In fig. \ref{fig:element}(a), we show pseudorapidity distributions
of charged particles, compared to data from ALICE \citet{alice} .
The two scenarios do not differ very much, and agree roughly with
the data. Also the multiplicity distribution agrees reasonably well
with data, see fig. \ref{fig:element}(b). We then investigate transverse
momentum distributions in fig. \ref{fig:element}(c). Here the base
calculation (without hydro) underestimates the data at intermediate
$p_{t}$ by a large factor, whereas the full calculation gets close
to the data. This is a very typical behavior of collective flow: the
distributions get harder at intermediate values of $p_{t}$ (around
1-5 GeV/$c$).

Experimentally, the ridge correlation is only observed for high multiplicity
events, and the effect is biggest for intermediate values of $p_{t}$
(1-3 GeV/c), and disappears towards large and small values. Why is
this so? We recall that also in pp the core-corona procedure is very
important: only regions with strongly overlapping strings contribute
to the core (and are treated via hydrodynamics), and this overlap
is more likely to happen in high multiplicity events. As a consequence
of the reduced hydro-contribution, the difference between the full
calculation and the {}``no hydro'' version is relatively small in
low multiplicity events (with multiplicities close to or smaller than
minimum bias), and therefore {}``collective effects'' like elliptical
flow or this ridge correlations will disappear with decreasing multiplicity.
The role of the transverse momenta can be seen from fig. \ref{fig:element}(c).
The main {}``flow effect'' appears at intermediate values of $p_{t}$
(1-3 GeV/c), as can be seen from the difference between the full calculation
and the one without hydro: particle production from a transversely
flowing liquid will produce preferentially intermediate $p_{t}$ particles.
At large $p_{t}$, there will be no effect, since these particles
originate from hard prcesses, not from the liquid. 

To summarize: our hydrodynamic approach based on flux tube initial
conditions, which has already been applied to explain very successfully
hundreds of of spectra in AuAu collisions at RHIC, and which excellently
describes the so-far published LHC spectra and Bose-Einstein correlation
functions, provides in a natural fashion a so-called near-side ridge
correlation in $\Delta\eta$ and $\Delta\phi$. This structure appears
as a consequence of a longitudinal invariant asymmetry of the energy
density from overlapping flux tubes, which translates into longitudinal
invariant elliptical flow.

\end{document}